\documentclass[ reprint,
amsmath,amssymb,
aps,
]{revtex4-2}

\usepackage{epsf,epsfig}
\usepackage{amssymb,amsmath,amsfonts}
\usepackage{color}
\usepackage{braket}
\usepackage{subcaption}
\usepackage[T1]{fontenc} 

\usepackage[dvipsnames, svgnames,  x11names ]{xcolor}
\usepackage[colorlinks]{hyperref}
\hypersetup{
   citecolor = {blue},
   linkcolor = {RubineRed},
}

\usepackage{physics}
\usepackage{tikz}
\usetikzlibrary{arrows,snakes,backgrounds}

\usepackage[english]{babel}

\usepackage{graphicx}

\definecolor{NordCyan}{HTML}{8FBCBB}          
\definecolor{NordBrightCyan}{HTML}{88C0D0}    
\definecolor{NordBlue}{HTML}{81A1C1}          
\definecolor{NordBrightBlue}{HTML}{5E81AC}    
\definecolor{NordRed}{HTML}{BF616A}           
\definecolor{NordOrange}{HTML}{D08770}        
\definecolor{NordYellow}{HTML}{EBCB8B}        
\definecolor{NordGreen}{HTML}{A3BE8C}         
\definecolor{NordMagenta}{HTML}{B48EAD}       

\begin{document}

\title{Modular Krylov Complexity as a Boundary Probe of  Area Operators and Entanglement Islands}


\author{Niloofar Vardian}
\email{vardian@sharif.edu}
\affiliation{Department of Physics, Sharif University of Technology,\\
P.O.Box 11155-9161, Tehran, Iran}
\affiliation{Research Center for High Energy Physics, Department of Physics, Sharif University of Technology,\\
P.O.Box 11155-9161, Tehran, Iran
}

\begin{abstract}
\noindent

We show that the area operator of a quantum extremal surface can be reconstructed directly from boundary dynamics without reference to bulk geometry. Our approach combines the operator-algebra quantum error-correction (OAQEC) structure of AdS/CFT with modular Krylov complexity. Using Lanczos coefficients of boundary modular dynamics, we extract the spectrum of the modular Hamiltonian restricted to the algebra of the entanglement wedge and isolate its central contribution, which is identified with the area operator. The construction is purely boundary-based and applies to superpositions of semiclassical geometries as well.  As a prospective application, we discuss island formation and the Page transition in evaporating black holes using boundary modular evolution, bypassing any bulk extremization. More broadly, our results establish modular Krylov complexity as a  probe of emergent spacetime geometry, providing a new route to accessing black hole interiors from boundary quantum dynamics.

\end{abstract}
\maketitle

\section{Introduction}

Understanding how bulk spacetime geometry emerges
from boundary quantum dynamics is a central
problem in holography. A precise connection is
provided by the Ryu--Takayanagi (RT) formula
\cite{Ryu:2006bv} and its quantum generalization,
which express boundary entanglement entropy
in terms of a bulk area contribution and bulk
entanglement entropy
\cite{Faulkner:2013ana,Engelhardt:2014gca}.
The resulting quantum extremal surface (QES)
prescription plays a central role in the
island description of the Page curve for
evaporating black holes
\cite{Penington:2019npb,Almheiri:2019psf,Almheiri:2019hni}.
However, the QES itself is determined through
a gravitational extremization in the bulk.
This raises a fundamental question: Can geometric data such as the QES area be inferred directly from boundary observables, without any bulk extremization?

The quantum error-correction (QEC) structure
of AdS/CFT has clarified the relationship
between bulk geometry and boundary quantum
information \cite{almheiri2015bulk}.
The entanglement wedge provides the bulk
region associated with a boundary subsystem
\cite{Czech:2012bh,Wall:2012uf,Headrick:2014cta},
and operator-algebra quantum error correction
(OAQEC) gives a precise formulation of its
reconstruction
\cite{dong2016reconstruction,Harlow:2016vwg}.
In this formulation, the area term is a central
operator in the protected bulk algebra
\cite{Harlow:2016vwg}. The JLMS relation connects
this operator to the difference between
boundary and bulk modular Hamiltonians
\cite{Jafferis:2015lma}.
They also motivate a
reconstruction question: how can the logical
modular generator, and hence the central
area contribution, be extracted using
boundary modular data and a specified
representation of the protected algebra?

Moreover, entanglement entropy alone cannot
capture the continued growth of black hole
interiors or all aspects of the long-time
dynamics of holographic states. In relevant
settings, entanglement can saturate while
interior growth continues
\cite{Susskind2014EntanglementNotEnough,StanfordSusskind2014}.
To address this limitation, quantum
computational complexity has been proposed
as an additional boundary quantity related
to bulk features such as interior volume
or action
\cite{Susskind2014Complexity,BrownSusskind2016,BrownSusskind2016b}.
This idea has motivated extensive work on
holographic complexity
\cite{BrownSusskind2016,BrownSusskind2016b,
couch2018holographic,barbon2015holographic,
Alishahiha:2017hwg,Alishahiha:2018tep,
Alishahiha:2018lfv,Akhavan:2018wla,
Belin:2021bga,Belin:2022xmt}.
However, the complexity = volume and
complexity = action proposals do not by
themselves provide a general procedure for
reconstructing spacetime from boundary
gates, circuits, or operator growth.
The observation that entanglement is not
enough therefore motivates additional
boundary diagnostics, while a controlled
reconstruction requires specifying the
information and assumptions on which
it depends.

In this work, we address the area-reconstruction
question. 
We combine OAQEC with the operator-space
Krylov framework underlying modular Krylov
complexity \cite{Caputa:2023vyr}.
Krylov subspace methods originate in numerical
linear algebra, where algorithms such as
Lanczos iteration provide reduced
representations of large linear operators
\cite{Lanczos1950}. Their application to
operator dynamics also underlies Krylov
complexity \cite{Parker:2018yvk}.
Recently, the study of Krylov complexity has been extended to holographic contexts \cite{Rabinovici:2023yex, Caputa:2024sux, Nongmaithem:2026rop, Fatemiabhari:2026eem, Bhattacharyya:2026zpu, Bhattacharya:2026ybq, Chatzis:2026ekd}.
Here we use this framework for reconstruction:
sufficiently complete operator-resolved modular
Krylov data determine the logical modular
generator within each factor sector up to
an additive constant. Density-matrix
normalization and sector probabilities obtained independently of the commutator data, for example through boundary central-projector measurements fix the remaining
constants. Subtracting the resulting
algebraic modular Hamiltonian from the
absolute compressed boundary modular
Hamiltonian then isolates the area operator.

The contributions of this work are as follows:
\begin{itemize}

    \item
    \textbf{An explicit extraction of the area
    operator.}
    Combining the reconstructed logical
    generators with normalization, sector
    probabilities, and the absolute boundary
    modular Hamiltonian gives an explicit
    procedure for recovering the central
    area operator. An exact tensor-network
    code provides a consistency benchmark.

    \item
    \textbf{A conditional application to islands
    and the Page transition.}
    For an evaporating black hole coupled to
    a bath, we discuss how the modular
    Hamiltonian of the entire CFT could be
    used to reconstruct a time-dependent
    area term, provided an appropriate
    protected algebra and the required
    modular-flow conditions are established.
    Identifying this term with the island
    boundary requires an additional
    complementary-wedge assumption.

\end{itemize}

More precisely, our construction starts from
the boundary modular Hamiltonian and uses
the Lanczos algorithm to reconstruct its
induced commutator on a specified protected
algebra, without prior knowledge of the
bulk modular Hamiltonian. Within each
factor sector, sufficiently complete
operator-resolved Krylov data determine
the logical generator up to a scalar,
which is fixed by density-matrix
normalization. Comparing this reconstructed
generator with the absolute boundary
modular Hamiltonian represented on the
code then isolates the central area term.
When fixed-area code-preservation or
protected modular-flow intertwining
conditions hold, the required dynamics
can be accessed directly through boundary
evolution without intermediate code
projections. This provides a dynamical
implementation of the OAQEC reconstruction
rather than requiring direct determination
of its ancillary states. The method
extends to multiple semiclassical sectors
with independently supplied sector
probabilities and requires no additional
bulk extremization within the specified
code. Applications to time-dependent
geometries require these assumptions
to be justified at each physical time.

\section{Area operator}
\label{area-operator}

The RT formula relates the
leading semiclassical entanglement entropy of
a boundary region $A$ to the area of a bulk
minimal surface homologous to $A$
\cite{Ryu:2006bv}. Its quantum generalization
involves the generalized entropy,
\begin{equation}
    S(\rho_A)
    =
    \frac{\operatorname{Area}(X_A)}{4G_N}
    +
    S_{\mathrm{bulk}}(\rho_a)
    +\cdots,
    \label{eq:generalized_entropy_area}
\end{equation}
evaluated on the appropriate dominant QES $X_A$. Here $a$ denotes
the entanglement wedge of $A$, and the ellipsis
denotes corrections beyond the approximation
being used.
The entanglement wedge of a boundary region \(A\) is the bulk domain of dependence bounded by \(A\) and its associated quantum extremal surface, encompassing the bulk degrees of freedom that can be reconstructed from the boundary subregion \(A\).
The area and bulk entropy terms
must be defined with a consistent regulator
and renormalization prescription.

Within a suitable holographic code subspace,
the JLMS relation expresses the corresponding
connection between boundary and bulk modular
Hamiltonians \cite{Jafferis:2015lma}. Let
$P_{\mathrm{code}}$ project onto the encoded
code subspace, and let
\begin{equation}
    K_A=-\log\rho_A,
    \qquad
    K_a=-\log\rho_a.
\end{equation}
For a single-factor code, the modular relation
takes the form
\begin{equation}
    \left.
    P_{\mathrm{code}}
    (K_A\otimes I_{\bar A})
    P_{\mathrm{code}}
    \right|_{\mathcal H_{\mathrm{code}}}
    =
    K_a\otimes I_{\bar a}
    +
    \mathcal L_A.
    \label{JLMSORD}
\end{equation}
Logical operators on the right-hand side are
identified with their encoded action on the
code. In a gravitational interpretation,
$\mathcal L_A$ is the QES area operator in
units of $4G_N$, at the order where the
Einstein-gravity area term applies.

OAQEC provides a precise finite-dimensional
setting for this structure
\cite{dong2016reconstruction,Harlow:2016vwg}.
The protected bulk observables form an algebra
$\mathcal A$ on $\mathcal H_{\mathrm{code}}$.
Complementary recovery means that operators
in $\mathcal A$ admit representatives on
the boundary region $A$, while operators
in its commutant admit representatives on
$\bar A$. Under these conditions, the
boundary entropy obeys an algebraic RT
relation,
\begin{equation}
    S(\rho_A)
    =
    \operatorname{Tr}_{\mathrm{code}}
    (\rho_{\mathrm{code}}\mathcal L_A)
    +
    S(\rho_{\mathrm{code}},\mathcal A),
    \label{eq:main_algebraic_rt}
\end{equation}
where $S(\rho_{\mathrm{code}},\mathcal A)$
is the entropy of the state restricted to
the protected algebra. The operator
$\mathcal L_A$ belongs to its center,
\begin{equation}
    \mathcal L_A\in Z(\mathcal A),
    \qquad
    [\mathcal L_A,O]=0
    \quad
    \forall  O\in\mathcal A.
    \label{eq:main_area_centrality}
\end{equation}
Centrality refers to the protected algebra,
not to the full algebra of boundary
operators on $A$.

In finite dimensions, the protected algebra
admits a decomposition
\begin{align}
    \mathcal H_{\mathrm{code}}
    &=
    \bigoplus_\alpha
    \left(
        \mathcal H_{a_\alpha}
        \otimes\mathcal H_{\bar a_\alpha}
    \right),
    \\
    \mathcal A
    &=
    \bigoplus_\alpha
    \left(
        \mathcal B(\mathcal H_{a_\alpha})
        \otimes I_{\bar a_\alpha}
    \right).
    \label{eq:main_algebra_decomposition}
\end{align}
The central projectors $P_\alpha$ label
the sectors of this decomposition. In
the semiclassical interpretation considered
here, these sectors can distinguish
different geometries or area sectors.
Within each factor block, a central
operator is proportional to the identity.
Consequently,
\begin{equation}
    \mathcal L_A
    =
    \sum_\alpha \ell_\alpha P_\alpha,
    \qquad
    \ell_\alpha
    =
    \frac{\operatorname{Area}(X_A^\alpha)}{4G_N},
    \label{eq:main_sector_area}
\end{equation}
when the geometric identification is
applicable.
For a single block, Eq.~\eqref{eq:main_sector_area}
reduces to a scalar operator on the code.
For several blocks, $\mathcal L_A$ remains
central but need not be proportional to
the identity on the full code space.
The block decomposition also accommodates
states with support on several sectors,
including coherent superpositions in the
underlying code Hilbert space  \cite{Almheiri:2016blp}.

To specify the state restricted to the
protected algebra, define the sector
probabilities and normalized conditional
logical states by
\begin{align}
    p_\alpha
    &=
    \operatorname{Tr}_{\mathrm{code}}
    (\rho_{\mathrm{code}}P_\alpha),
    \\
    \rho_{a_\alpha}
    &=
    \frac{1}{p_\alpha}
    \operatorname{tr}_{\bar a_\alpha}
    \left(
        P_\alpha\rho_{\mathrm{code}}P_\alpha
    \right),
    \qquad p_\alpha>0.
\end{align}
The projected operator is restricted to
the $\alpha$ block before taking the
partial trace. Consequently,
$\operatorname{tr}_{a_\alpha}\rho_{a_\alpha}=1$.

On the occupied support, and using the
direct-sum trace over the logical factors,
its representation on the code is
\begin{equation}
    K\Big|_{\mathcal A}
    =
    \bigoplus_\alpha
    \left(
        -\log\rho_{a_\alpha}
        -
        \log p_\alpha\,I_{a_\alpha}
    \right)
    \otimes I_{\bar a_\alpha}.
    \label{eq:main_algebra_modular}
\end{equation}
The corresponding algebraic entropy is
\begin{equation}
    S(\rho_{\mathrm{code}},\mathcal A)
    =
    -\sum_\alpha p_\alpha\log p_\alpha
    +
    \sum_\alpha p_\alpha S(\rho_{a_\alpha}).
    \label{eq:main_algebra_entropy}
\end{equation}
The identities on the complementary
logical factors specify the representation;
they are not included in the trace used
to normalize each $\rho_{a_\alpha}$ \cite{Kamal:2019skn}.

The general code-space modular identity
is therefore
\begin{equation}
    \left.
    P_{\mathrm{code}}
    (K_A\otimes I_{\bar A})
    P_{\mathrm{code}}
    \right|_{\mathcal H_{\mathrm{code}}}
    =
    K\Big|_{\mathcal A}
    +
    \mathcal L_A.
    \label{JLMSJLMS}
\end{equation}
This identity is exact in finite-dimensional
complementary-recovery codes with the
corresponding algebraic RT structure.
Its interpretation in semiclassical
gravity is subject to the accuracy of
the effective code and the gravitational
modular relation.

In the canonical complementary-recovery
construction, the eigenvalues $\ell_\alpha$
are the entropies of fixed ancillary
states associated with the encoding
\cite{Harlow:2016vwg}. This gives an
explicit algebraic origin for the
central term. The origin of the central term in OAQEC is reviewed in Appendix \ref{App1}.
Our objective is to recover this term
by combining boundary modular information
with a reconstruction of the normalized
logical modular generators. The modular-flow conditions
and data required by the reconstruction
are specified in
Secs.~ \ref{sec:area_modular_krylov-data} and \ref{sec:multisector_reconstruction}.

\section{Krylov subspace methods}
\label{spectrum}

Krylov subspace methods originate in numerical
linear algebra, where they provide efficient
representations of large linear operators on
subspaces generated by repeated operator action.
They underpin widely used algorithms for
eigenvalue problems, linear systems, and matrix
functions. The Lanczos algorithm is the
Hermitian version of this construction
\cite{Lanczos1950,viswanath1994recursion}.
These methods are now employed throughout
scientific computing and data analysis, but
their underlying principle is independent of
any particular physical application.

For a Hermitian generator $G$ and a normalized
seed $|\psi_0\rangle$, define
\begin{equation}
    \mathcal K_m(G,\psi_0)
    =
    \operatorname{span}
    \left\{
        |\psi_0\rangle,
        G|\psi_0\rangle,
        \ldots,
        G^{m-1}|\psi_0\rangle
    \right\}.
    \label{eq:krylov_finite_subspace}
\end{equation}
Lanczos orthogonalization produces an
orthonormal basis $\{|K_n\rangle\}$ satisfying
\begin{equation}
    G|K_n\rangle
    =
    b_{n+1}|K_{n+1}\rangle
    +
    a_n|K_n\rangle
    +
    b_n|K_{n-1}\rangle,
    \label{eq:main_lanczos_recursion}
\end{equation}
with $b_0=0$. If $Q_m$ has the Krylov basis
vectors as its columns, the projected matrix is
\begin{equation}
    T_m=Q_m^\dagger GQ_m
    =
    \begin{pmatrix}
        a_0 & b_1 & 0 & \cdots\\
        b_1 & a_1 & b_2 & \cdots\\
        0 & b_2 & a_2 & \ddots\\
        \vdots & \vdots & \ddots & \ddots
    \end{pmatrix}.
    \label{eq:main_krylov_matrix}
\end{equation}
When
the recursion terminates, the generated
subspace is invariant and $T_m$ represents
the exact restriction of $G$ to that subspace.

The information accessible to a single seed
is characterized by its spectral measure.
Writing
\begin{equation}
    G=\sum_\lambda \lambda P_\lambda,
\end{equation}
where $P_\lambda$ projects onto the eigenspace
with eigenvalue $\lambda$, the complete
seed-generated Krylov subspace is
\begin{equation}
    \mathcal K(G,\psi_0)
    =
    \operatorname{span}
    \left\{
        P_\lambda|\psi_0\rangle:
        P_\lambda|\psi_0\rangle\neq0
    \right\}.
    \label{eq:main_krylov_spectral_support}
\end{equation}
A single seed therefore accesses only its
supported eigenvalues and at most one
direction within each degenerate eigenspace.

The Lanczos coefficients need not be obtained
by explicitly applying a known matrix
representation of the generator. They can
also be reconstructed from moments of a
return amplitude (see Appendix \ref{algorithm}). This determines the
seed-dependent spectral measure, but does
not by itself determine the generator in
a specified logical basis. The reconstruction
below additionally requires sufficient
operator-resolved information.



Krylov projection has also been adapted to
quantum algorithms like quantum Krylov
diagonalization (KQD) \cite{Cortes:2021qkd,Stair:2022rqkd}.
KQD and Lanczos reconstruction share the
principle of replacing a large spectral
problem by one defined on an accessible
subspace. Their implementations and data
requirements are nevertheless different.
In particular, the present work does not
require a quantum device or claim a quantum
computational advantage. We use the
classical operator-space Lanczos formulation,
with modular evolution supplying the
relevant dynamical data.

\section{Area reconstruction from modular Krylov data:
single sector}
\label{sec:area_modular_krylov-data}

Our starting point is the boundary modular
Hamiltonian $K_A=-\log\rho_A$, which we assume
is available. We do not assume knowledge of
the bulk modular Hamiltonian. Instead, we
use the action of $K_A$ on protected boundary
operators to reconstruct the logical modular
generator. The central contribution is
invisible to this operator evolution and
is recovered by comparing the reconstructed
generator with the absolute boundary modular
Hamiltonian on the code.

We first consider a single regulated sector,
with
\begin{equation}
    \mathcal H_{\mathrm{code}}
    =
    \mathcal H_a\otimes\mathcal H_{\bar a},
    \qquad
    \mathcal A
    =
    \mathcal B(\mathcal H_a)\otimes I_{\bar a}.
    \label{eq:single_factor_algebra}
\end{equation}
The protected algebra is a factor, so its
central area operator is proportional to
the identity. We assume that the protected
boundary representatives and their
identification with this logical algebra
are specified. This is structural information
about the code, not prior knowledge of the
logical modular Hamiltonian.

Let $V$ denote the encoding isometry and
define the boundary modular generator
represented on the code by
\begin{equation}
    K_A^{\mathrm{code}}
    =
    V^\dagger
    (K_A\otimes I_{\bar A})V.
    \label{eq:single_boundary_code_generator}
\end{equation}
The single-sector modular identity is
\begin{equation}
    K_A^{\mathrm{code}}
    =
    K_a\otimes I_{\bar a}
    +
    \ell_A I_{\mathrm{code}},
    \label{eq:single_modular_decomposition}
\end{equation}
where $K_a=-\log\rho_a$ is unknown and
$\operatorname{tr}_a\rho_a=1$.
Equivalently, using the sector projector
$P= P_{code}$ and suppressing the spectator identity,
we write
$K_a=P K|_{\mathcal A}P$, understood as
a restriction to the occupied sector.
Equation~\eqref{eq:single_modular_decomposition}
specifies the decomposition to be
reconstructed; it does not supply $K_a$
as an input.
Moreover, we work in regulated finite dimensions and assume that each occupied conditional logical state is faithful on the factor being reconstructed. Otherwise, all modular operators and reconstruction statements are restricted to the appropriate support.

\subsection{Boundary modular evolution and the Krylov representation}

For a protected boundary representative
$O_A$, consider modular evolution
\begin{equation}
    O_A(s)
    =
    e^{isK_A}O_Ae^{-isK_A}.
\end{equation}
We assume that this evolution intertwines
with the logical modular action on the
chosen protected operators. In the logical
representation, its generator is therefore
\begin{equation}
    \mathcal L_S(O)
    =
    [K_a,O],
    \qquad
    O\in\mathcal B(\mathcal H_a).
    \label{eq:single_induced_derivation}
\end{equation}
The intertwining is an additional property
of the protected representatives, not a
consequence of compression alone.
It allows the action in
Eq.~\eqref{eq:single_induced_derivation}
to be obtained from boundary modular
evolution without knowing $K_a$.

The reason that this dynamics isolates
the noncentral contribution is immediate:
\begin{equation}
    [K_a\otimes I_{\bar a}
        +\ell_A I_{\mathrm{code}},
      O\otimes I_{\bar a}]
    =
    [K_a,O]\otimes I_{\bar a}.
\end{equation}
Thus the unknown area contribution does
not enter the protected commutator.

Choose a normalized protected seed $O_0$
and use the logical Hilbert--Schmidt
inner product
\begin{equation}
    (O_1,O_2)
    =
    \operatorname{tr}_a(O_1^\dagger O_2).
\end{equation}
Lanczos iteration applied to the induced
boundary modular action produces
\begin{equation}
    \mathcal L_S O_n
    = a_n O_n+
    b_{n+1}O_{n+1}
    +
    b_nO_{n-1},
    \label{eq:single_operator_lanczos}
\end{equation}
with $b_0=0$. Let $Q_m$ contain the
coordinates of the first $m$ Krylov
operators in a known orthonormal logical
operator basis. The Lanczos matrix is
\begin{equation}
    T_m=Q_m^\dagger\mathcal L_S Q_m.
    \label{eq:single_lanczos_projection}
\end{equation}
It represents the modular commutator
on the selected operator subspace,
rather than the modular Hamiltonian
as an operator on states.

At termination, this subspace is invariant,
and $T_m$ determines the exact restricted
derivation.
An arbitrary truncation,
however, need not determine its action
on the entire protected algebra.

For any orthonormal Krylov basis matrix $Q_m$, let
$P_m = Q_m Q_m^\dagger$ denote the orthogonal projector onto
the corresponding subspace of the logical operator Hilbert space.
Since $T_m = Q_m^\dagger \mathcal{L}_S Q_m$, we have
\[
Q_m T_m Q_m^\dagger
=
P_m \mathcal{L}_S P_m.
\]
At exact Lanczos termination, this subspace is invariant under
$\mathcal{L}_S$, and the right-hand side equals
$\mathcal{L}_S P_m$. Thus, the Krylov data determine the exact
action of the derivation on the seed-generated subspace, but do not
generally determine its action on the full logical operator space.
To reconstruct the full derivation, we seek a protected operator whose Krylov subspace has the appropriate dimension. 
In other word the operator should be such that all
transition frequencies $\lambda_i - \lambda_j$ are distinct and $O_0$ has non-vanishing matrix elements between all corresponding eigenstates.
If a chosen seed is insufficient, we replace it with another and test the identifiability condition again. Denoting the resulting
complete basis matrix by $Q$ and setting
$T = Q^\dagger \mathcal{L}_S Q$, we have
$Q Q^\dagger = I_l$, where $I_l$ is the
identity on the logical operator Hilbert space, and hence
\begin{equation}
\mathcal{L}_S = Q T Q^\dagger.
\label{eq:full_derivation_reconstruction}
\end{equation}

For the reconstruction below we require
sufficient operator-resolved data to
identify the derivation, obtained, for
example, from appropriate seeds whose Krylov spaces span the logical
operator space. Retaining the coordinates
of the Krylov operators is essential;
scalar Lanczos coefficients alone do
not specify the derivation in a known
logical basis.

t is worth noting that while different choices of cyclic seed operators
generate different orthonormal Krylov bases—and therefore yield different tridiagonal matrix elements for $\mathcal{L}_S$ all such tridiagonal representations are related by unitary similarity transformations. Diagonalizing any of them yields the identical invariant spectrum of the Liouvillian in its eigenbasis, guaranteeing that the reconstructed generator spectrum and the extracted are independent of the choice of initial seed.

\subsection{Recovering the unknown logical generator}

The next step converts the reconstructed
derivation into a generator acting on
the logical Hilbert space. Choose the
vectorization convention
\begin{equation}
    |i\rangle\langle j|
    \longmapsto |i\rangle\otimes|j\rangle.
\end{equation}
In this representation,
\begin{equation}
    \mathcal L_S
    =
    K_a\otimes I_a
    -
    I_a\otimes K_a^T.
    \label{eq:single_vectorized_derivation}
\end{equation}
The basis is a specified logical basis;
it need not be an eigenbasis of the
unknown $K_a$.

Writing $d_a=\dim\mathcal H_a$, define
the generator recovered from the
boundary-induced Krylov data by
\begin{equation}
\begin{split}
    K_{\mathrm{rec}}
    &=
    \frac{1}{d_a}
    \operatorname{Tr}_2(QTQ^\dagger)
    \\
    &=
    K_a-
    \frac{\operatorname{tr}_a K_a}{d_a}I_a.
\end{split}
    \label{eq:single_reconstructed_generator}
\end{equation}

In the following partial-trace formula, $Q$ is understood to contain
the coordinates of the assembled operator basis in the matrix-unit
basis $\{|i\rangle\langle j|\}$.
Thus $QTQ^\dagger$ represents the full derivation on the vectorized
logical operator space, and $\operatorname{Tr}_2$ denotes the
ordinary, unnormalized partial trace over its second tensor factor.

This reconstructs the unknown logical
generator up to an additive scalar.
The scalar is fixed by the normalization
of the logical density matrix.
Set
\begin{equation}
    Z_{\mathrm{rec}}
    =
    \operatorname{tr}_a
    e^{-K_{\mathrm{rec}}}.
\end{equation}
The reconstructed normalized modular
Hamiltonian is
\begin{equation}
    K_a^{\mathrm{rec}}
    =
    K_{\mathrm{rec}}
    +
    \log Z_{\mathrm{rec}}\,I_a.
    \label{eq:single_normalized_reconstruction}
\end{equation}
It satisfies
$\operatorname{tr}_a e^{-K_a^{\mathrm{rec}}}=1$
and equals $K_a$ for exact data under
the stated assumptions. No prior
matrix elements or eigenvalues of
$K_a$ are used in this step.

 One  important point  is  that  the  Krylov  space  is  blind  to degeneracies  in  the  Hilbert  space
In this case, the boundary and logical modular Hamiltonians differ by a number times the identity. Therefore, they have the same degeneracies in the logical factor. The Krylov subspace may not give the whole logical spectrum, but if we have enough information to fix its normalization, we can obtain the area coefficient from the difference between the two modular Hamiltonians.

\subsection{Recovering the area by subtraction}

We now compare the reconstructed logical
generator with the known absolute
boundary modular generator represented
on the same code space:
\begin{equation}
    \mathcal L_A^{\mathrm{rec}}
    =
    K_A^{\mathrm{code}}
    -
    K_a^{\mathrm{rec}}\otimes I_{\bar a}.
    \label{eq:single_central_subtraction}
\end{equation}
For an exact single-sector code,
\begin{equation}
    \mathcal L_A^{\mathrm{rec}}
    =
    \ell_A I_{\mathrm{code}}.
\end{equation}
When the geometric interpretation applies,
$\ell_A$ is the area contribution
associated with the specified RT or
quantum extremal surface, in units
of $4G_N$, with the appropriate
coarse-graining and renormalization
conventions.

The subtraction is between state-space
operators expressed in the same logical
representation. It is not a subtraction
of the operator-space Lanczos matrix
from $K_A$. Representing both quantities
only through their commutators would
erase the central contribution from
both and could not recover the area.
The absolute boundary modular input
must therefore be retained in
Eq.~\eqref{eq:single_central_subtraction}.

The role of the Krylov construction is
to obtain the unknown logical generator
from the dynamics induced by the known
boundary generator. Knowledge of the
bulk modular Hamiltonian is not an
input. What is required is a specified
protected factor algebra, sufficiently
informative operator-resolved modular
data, and access to the absolute
boundary modular operator in the
corresponding code representation.
An explicit microscopic encoding need
not be constructed if these data and
identifications are independently
available.

Thus the central term, although absent
from protected modular dynamics, is
recovered as the difference between
the absolute boundary generator and
the normalized logical generator
reconstructed from those dynamics.
The procedure determines the area
term in the specified code; it does
not independently determine the
surface's location or its dominance
among competing gravitational saddles.

\section{Area reconstruction from modular Krylov data:
multiple sectors}
\label{sec:multisector_reconstruction}

The single-sector construction reconstructs
the unknown logical modular generator from
the action of the known boundary modular
Hamiltonian on protected operators. We now
extend this procedure to a code containing
several semiclassical sectors. A central
question is whether these dynamics can
be accessed through boundary modular
evolution without explicitly inserting
the code projector.

\subsection{Unprojected boundary modular flow}

The OAQEC modular identity is initially
an identity between compressed generators:
\begin{equation}
    P_{\mathrm{code}}
    (K_A\otimes I_{\bar A})
    P_{\mathrm{code}}
    =
    V\left(
        K\big|_{\mathcal A}+\mathcal L_A
    \right)V^\dagger,
    \label{eq:multi_compressed_identity}
\end{equation}
where the right-hand side is understood
as an operator supported on the encoded
code. Compression alone does not imply
that boundary modular evolution preserves
this subspace.

Additional bulk calculations provide
more information in fixed-area settings
\cite{Dong:2018seb,Akers:2018fow}.
At the semiclassical order considered
in those constructions, the fixed-area
contribution to the entanglement spectrum
is flat \cite{Dong:2018seb,Akers:2018fow,Dong:2019piw}. 
In this exact canonical setting,
the encoded code is invariant under
the boundary modular generator.
On the relevant support,
\begin{equation}
    [K_A\otimes I_{\bar A},
        P_{\mathrm{code}}]=0.
    \label{eq:multi_code_commutation}
\end{equation}

The gravitational use of this argument
is therefore restricted to the order
and class of states for which the
fixed-area and modular-flow results
apply. The large-code analysis of
~\cite{Dong:2026ysd}
clarifies additional conditions needed
to control exponentiated JLMS across
semiclassical sectors. In particular,
control for selected states must be
distinguished from uniform control
throughout a code.

When code preservation holds to the
required accuracy, one may omit the
explicit projections when the modular
identity acts on an encoded code state:
\begin{equation}
    (K_A\otimes I_{\bar A})V
    \simeq
    V\left(
        K\big|_{\mathcal A}+\mathcal L_A
    \right).
    \label{eq:multi_uncompressed_code_action}
\end{equation}
This is an equality, or controlled
approximation, on the encoded code;
it is not an identity on the full
boundary Hilbert space. It permits
boundary modular evolution to generate
the required logical dynamics without
repeated projector insertions.
Even when preservation of all code
vectors is unavailable, intertwining
on the chosen protected operators
can suffice for the dynamical part
of the reconstruction.

\subsection{Sectorwise reconstruction}

Let
\begin{align}
    \mathcal H_{\mathrm{code}}
    &=
    \bigoplus_\alpha
    \left(
        \mathcal H_{a_\alpha}
        \otimes\mathcal H_{\bar a_\alpha}
    \right),
    \\
    \mathcal A
    &=
    \bigoplus_\alpha
    \left[
        \mathcal B(\mathcal H_{a_\alpha})
        \otimes I_{\bar a_\alpha}
    \right].
    \label{eq:multi_protected_algebra}
\end{align}
Each regulated sector is a factor,
while the direct-sum algebra has
central projectors $P_\alpha$.
The sector labels can distinguish
area windows and semiclassical
background data; they need not specify
unique complete geometries.

For an occupied sector, write
\begin{equation}
    p_\alpha
    =
    \operatorname{Tr}_{\mathrm{code}}
    (\rho_{\mathrm{code}}P_\alpha),
    \qquad
    K_{a_\alpha}
    =
    -\log\rho_{a_\alpha},
\end{equation}
with
$\operatorname{tr}_{a_\alpha}
\rho_{a_\alpha}=1$.
The corresponding block of the
algebraic modular Hamiltonian is
\begin{equation}
    \left.
    P_\alpha K\big|_{\mathcal A}P_\alpha
    \right|_{\mathcal H_\alpha}
    =
    \left(
        K_{a_\alpha}
        -\log p_\alpha\,I_{a_\alpha}
    \right)\otimes I_{\bar a_\alpha}.
    \label{eq:multi_modular_block}
\end{equation}
Thus the conditional generator
$K_{a_\alpha}$ and the full sector
block differ by the probability term.
They nevertheless induce the same
commutator within the factor.

Under the assumed modular intertwining,
boundary evolution of protected
sector operators determines
\begin{equation}
    \mathcal L_{S,\alpha}(O)
    =
    [K_{a_\alpha},O],
    \qquad
    O\in\mathcal B(\mathcal H_{a_\alpha}).
    \label{eq:multi_sector_derivation}
\end{equation}
Neither $-\log p_\alpha$ nor the
central area coefficient appears
in this dynamics.

Apply the single-sector reconstruction
independently to each factor. With
sufficient operator-resolved Krylov
data, let
\begin{equation}
    \mathcal L_{S,\alpha}
    =
    Q_\alpha T_\alpha Q_\alpha^\dagger
\end{equation}
denote the reconstructed derivation
in a known logical operator basis.
As in the single-sector case, this
requires sufficient information to
identify the derivation, not an
arbitrary scalar Lanczos truncation.

Using the same vectorization convention,
\begin{equation}
    K_{\mathrm{rec},\alpha}
    =
    \frac{1}{d_\alpha}
    \operatorname{Tr}_2
    \left(
        Q_\alpha T_\alpha Q_\alpha^\dagger
    \right),
    \qquad
    d_\alpha=\dim\mathcal H_{a_\alpha}.
\end{equation}
Conditional-state normalization gives
\begin{equation}
    K_{a_\alpha}^{\mathrm{rec}}
    =
    K_{\mathrm{rec},\alpha}
    +
    \log\!\left(
        \operatorname{tr}_{a_\alpha}
        e^{-K_{\mathrm{rec},\alpha}}
    \right)I_{a_\alpha}.
    \label{eq:multi_normalized_generator}
\end{equation}
Thus no bulk modular Hamiltonian is
supplied as an input: each conditional
generator is reconstructed from the
boundary-induced operator dynamics.

\subsection{Central subtraction across sectors}

The sector probabilities are not determined by within-sector modular commutator dynamics. They can, however, be obtained from additional boundary measurements. In the specified complementary-recovery code, each central projector admits a representative on region A. Choosing these representatives as orthogonal projectors on the relevant boundary support, their expectation values in the boundary state give the sector probabilities. Thus, the probabilities required below are independent of the commutator data, but need not constitute additional bulk input.

Given these probabilities, assemble
\begin{equation}
    K\big|_{\mathcal A}^{\mathrm{rec}}
    =
    \bigoplus_\alpha
    \left(
        K_{a_\alpha}^{\mathrm{rec}}
        -\log p_\alpha\,I_{a_\alpha}
    \right)\otimes I_{\bar a_\alpha}.
    \label{eq:multi_assembled_generator}
\end{equation}
The absolute boundary modular generator
in the same code representation is
\begin{equation}
    K_A^{\mathrm{code}}
    =
    V^\dagger
    (K_A\otimes I_{\bar A})V.
\end{equation}
The reconstructed area operator is
\begin{equation}
    \mathcal L_A^{\mathrm{rec}}
    =
    K_A^{\mathrm{code}}
    -
    K\big|_{\mathcal A}^{\mathrm{rec}}.
    \label{eq:multi_area_subtraction}
\end{equation}
In an exact code,
\begin{equation}
    \mathcal L_A^{\mathrm{rec}}
    =
    \sum_\alpha\ell_\alpha P_\alpha.
\end{equation}
The result is scalar in each factor,
but need not be scalar on the whole
code.

The geometric interpretation of
$\ell_\alpha$ depends on the code
and its coarse-graining prescription.
In particular, the large-code
construction of
~\cite{Dong:2026ysd}
involves a coarse-grained area with
contributions associated with the
area windows. The subtraction recovers
the central operator appearing in
the specified modular identity;
its identification with a geometric
area requires consistent conventions.

\subsection{Role of the dynamical construction}

The unprojected boundary modular dynamics argument explains a
potential practical advantage of the
present implementation of OAQEC, especially in a single-sector case.
If the relevant code-preservation
or protected-flow conditions are
established, the logical modular
dynamics can be accessed directly
through boundary evolution, without
implementing the code projector at
every step. The Krylov procedure then
reconstructs the unknown logical
generators rather than requiring
their prior determination or an
explicit reconstruction of the
ancillary density matrices.

This is an advantage in the organization
and possible acquisition of the
required data, not a replacement
for the OAQEC framework. The protected
algebra, its boundary representatives,
and their logical identification
remain necessary. Likewise, the
subtraction requires absolute boundary
modular information in that logical
representation. Dropping projector
insertions during evolution does
not eliminate these requirements.

\section{An exact tensor-network benchmark}
\label{sec:tn_benchmark}

We now test the reconstruction in an explicit
single-sector code whose area operator is known
independently. The example realizes the elementary
entangled-bond contribution to a tensor-network RT formula. It permits an analytic
calculation of the modular evolution, Lanczos
coefficients, normalized logical modular Hamiltonian,
and reconstructed area operator.

\subsection{Tensor-network interpretation and code}
\label{subsec:tn_code}

Tensor networks provide concrete settings in which
boundary entanglement can be related to the geometry
of an auxiliary network. In holographic
quantum-error-correcting tensor networks, logical
degrees of freedom are encoded into boundary degrees
of freedom, and appropriate network cuts reproduce
RT-type entropy relations
\cite{Pastawski2015}. Related minimal-cut formulas
arise in random tensor networks in the large-bond-dimension
regime \cite{Hayden2016}. A maximally entangled bond
of dimension $D$ contributes $\log D$ to the cut
entropy, provided the maps on either side preserve
the relevant entanglement. For a general tensor
network, bond counting alone is only an entropy
bound; saturation requires additional conditions.

The operator-algebraic origin of the corresponding
area term is particularly transparent in codes with
complementary recovery. In their canonical
decomposition, a fixed ancillary entangled state
contributes a central term to the boundary entropy
and modular Hamiltonian. Its entropy is the
eigenvalue of the area operator in the associated
sector \cite{Harlow:2016vwg}.

Here we use the simplest exact realization of this
structure: two local encoding tensors connected by
one maximally entangled bond. This is a minimal
tensor-network code, not a model of a complete
asymptotically AdS geometry. Nevertheless, its
discrete RT term is exact and provides an
unambiguous benchmark for our reconstruction.

Let the logical Hilbert space be
\begin{equation}
    \mathcal H_{\mathrm{code}}
    =
    \mathcal H_a\otimes\mathcal H_{\bar a},
    \qquad
    \dim\mathcal H_a
    =
    \dim\mathcal H_{\bar a}
    =
    2.
    \label{eq:tn_logical_space}
\end{equation}
Introduce ancillary systems $r$ and $\bar r$ of
dimension $D$, prepared in
\begin{equation}
    |\chi_D\rangle_{r\bar r}
    =
    \frac{1}{\sqrt D}
    \sum_{m=1}^{D}
    |m\rangle_r|m\rangle_{\bar r}.
    \label{eq:tn_entangled_bond}
\end{equation}
The boundary bipartition is
\begin{equation}
    \mathcal H_A
    =
    \mathcal H_a\otimes\mathcal H_r,
    \qquad
    \mathcal H_{\bar A}
    =
    \mathcal H_{\bar a}\otimes\mathcal H_{\bar r}.
    \label{eq:tn_boundary_spaces}
\end{equation}
An encoding isometry $V$ is defined on a logical
basis by
\begin{equation}
    V|i,j\rangle
    =
    \frac{1}{\sqrt D}
    \sum_{m=1}^{D}
    |i,m\rangle_A
    |j,m\rangle_{\bar A}.
    \label{eq:tn_encoding_isometry}
\end{equation}
Unitaries on $A$ and $\bar A$ could be included
without changing the conclusions; we use the
canonical representation in
Eq.~\eqref{eq:tn_encoding_isometry}.

The protected algebra is a single factor,
\begin{equation}
    \mathcal A
    =
    \mathcal B(\mathcal H_a)\otimes I_{\bar a},
    \qquad
    Z(\mathcal A)=\mathbb C I_{\mathrm{code}}.
    \label{eq:tn_factor_algebra}
\end{equation}
Logical operators on $a$ have representatives
$O_a\otimes I_r$ on $A$, while logical operators
on $\bar a$ have representatives on $\bar A$.
The code therefore has exact complementary
recovery.

For any logical density matrix $\rho$, the reduced
boundary state is
\begin{equation}
    \rho_A
    =
    \rho_a\otimes\frac{I_r}{D},
    \qquad
    \rho_a=\operatorname{tr}_{\bar a}\rho.
    \label{eq:tn_boundary_density_matrix}
\end{equation}
It follows that
\begin{equation}
    S(\rho_A)
    =
    S(\rho_a)+\log D.
    \label{eq:tn_rt_entropy_relation}
\end{equation}
The first term is the entropy of the logical
degrees of freedom on the $A$ side of the cut.
The second is the contribution of the single
entangled bond. In the OAQEC area relation, this
identifies the known area operator as
\begin{equation}
    \mathcal{L}_A^{\mathrm{known}}
    =
    (\log D)I_{\mathrm{code}}.
    \label{eq:tn_known_area}
\end{equation}
Equation~\eqref{eq:tn_known_area} is the discrete
RT quantity that we seek to recover. Its
identification follows directly from the code
structure and does not use the Krylov calculation.

\subsection{Boundary modular evolution and Lanczos data}
\label{subsec:tn_lanczos}

Choose the logical reference state
\begin{equation}
    |\psi\rangle
    =
    \sqrt q\,|0,0\rangle
    +
    \sqrt{1-q}\,|1,1\rangle,
    \qquad
    0<q<\frac12.
    \label{eq:tn_reference_state}
\end{equation}
Then
\begin{equation}
    \rho_a
    =
    \begin{pmatrix}
        q&0\\
        0&1-q
    \end{pmatrix},
    \qquad
    K_a=-\log\rho_a.
    \label{eq:tn_logical_density_matrix}
\end{equation}
The parameter $q$ specifies the state used to
generate the benchmark data. In the reconstruction
below, it is recovered from the modular
commutators.

The boundary modular Hamiltonian is
\begin{equation}
    K_A
    =
    K_a\otimes I_r
    +
    (\log D)I_A.
    \label{eq:tn_boundary_modular_hamiltonian}
\end{equation}
Consequently, its action on protected operators
is independent of the bond contribution:
\begin{equation}
   \mathcal  L_S(O_a\otimes I_r)
    =
    [K_A,O_a\otimes I_r]
    =
    [K_a,O_a]\otimes I_r.
    \label{eq:tn_protected_derivation}
\end{equation}
The ancillary reduced state is exactly
maximally mixed, so this benchmark
realizes the flat-spectrum condition
without approximation. In particular,
\begin{equation}
    (K_A\otimes I_{\bar A})V
    =
    V\left(
        K_a\otimes I_{\bar a}
        +
        \log D\,I_{\mathrm{code}}
    \right).
    \label{eq:tn_exact_modular_intertwining}
\end{equation}
The encoded code is therefore invariant
under the boundary modular generator.
Since this generator is Hermitian,
\begin{equation}
    [K_A\otimes I_{\bar A},
        P_{\mathrm{code}}]=0,
    \qquad
    P_{\mathrm{code}}=VV^\dagger.
    \label{eq:tn_code_preservation}
\end{equation}
Protected modular evolution can thus
be generated directly on the boundary
without intermediate code projections.
This provides an exact realization of
the mechanism discussed in the preceding
section.

Use the normalized Hilbert--Schmidt inner product
\begin{equation}
    (X,Y)
    =
    \frac{1}{2D}
    \operatorname{Tr}_A(X^\dagger Y),
    \label{eq:tn_operator_inner_product}
\end{equation}
and choose the normalized Hermitian seed
\begin{equation}
    O_0=\sigma_x\otimes I_r.
    \label{eq:tn_seed}
\end{equation}
Writing
\begin{equation}
    \omega
    =
    \log\frac{1-q}{q}>0,
    \label{eq:tn_modular_frequency}
\end{equation}
the first commutator is
\begin{equation}
    \mathcal{L}_S(O_0)
    =
    i\omega\sigma_y\otimes I_r.
    \label{eq:tn_first_commutator}
\end{equation}
The Lanczos recursion therefore gives
\begin{equation}
    a_0=0,
    \qquad
    b_1=\omega,
    \qquad
    O_1=i\sigma_y\otimes I_r.
    \label{eq:tn_first_lanczos_step}
\end{equation}
Although $O_1$ is anti-Hermitian, it is a
normalized vector in the complex operator
Hilbert space. Its modular commutator is
\begin{equation}
    \mathcal{L}_S(O_1)=\omega O_0.
    \label{eq:tn_second_commutator}
\end{equation}
Thus
\begin{equation}
    a_1=0,
    \qquad
    b_2=0,
    \qquad
    T=
    \begin{pmatrix}
        0&\omega\\
        \omega&0
    \end{pmatrix}.
    \label{eq:tn_lanczos_matrix}
\end{equation}
The chain terminates after two Krylov vectors,
and the eigenvalues of $T$ are  $\pm\omega$.

For completeness, the evolved seed is
\begin{align}
    O_0(s)
    &=
    e^{is\mathcal{L}_S}O_0
    \nonumber\\
    &=
    \cos(\omega s)O_0
    +
    i\sin(\omega s)O_1.
    \label{eq:tn_seed_evolution}
\end{align}
The operator return amplitude and Krylov
complexity are consequently
\begin{equation}
    (O_0,O_0(s))=\cos(\omega s),
    \qquad
    C_K(s)=\sin^2(\omega s).
    \label{eq:tn_return_and_complexity}
\end{equation}

\subsection{Reconstruction and normalization of the generator}
\label{subsec:tn_generator_reconstruction}

To recover the logical generator, we retain the
Krylov operators as well as the scalar Lanczos
coefficients. Equations
\eqref{eq:tn_first_commutator} and
\eqref{eq:tn_second_commutator} give the labeled
operator relations
\begin{equation}
    [K_{\mathrm{rec}},\sigma_x]
    =
    i\omega\sigma_y,
    \qquad
    [K_{\mathrm{rec}},i\sigma_y]
    =
    \omega\sigma_x.
    \label{eq:tn_generator_constraints}
\end{equation}
A general Hermitian qubit generator has the form
\begin{equation}
    K_{\mathrm{rec}}
    =
    cI_a+h_x\sigma_x+h_y\sigma_y+h_z\sigma_z.
\end{equation}
The two commutator relations imply
\begin{equation}
    h_x=h_y=0,
    \qquad
    h_z=\frac{\omega}{2}.
\end{equation}
Choosing the traceless representative yields
\begin{equation}
    K_{\mathrm{rec}}
    =
    \frac{\omega}{2}\sigma_z.
    \label{eq:tn_traceless_generator}
\end{equation}

The two-dimensional Krylov subspace does not
span the complete qubit operator space.
Nevertheless, the known qubit algebra and the
two operator-valued commutator equations fix
the generator modulo its scalar. This is a
special feature of the present example, not
a general assertion that one scalar Lanczos
chain determines an arbitrary modular
Hamiltonian.

The remaining scalar is fixed by normalizing
the conditional logical state:
\begin{align}
    K_a^{\mathrm{rec}}
    &=
    K_{\mathrm{rec}}
    +
    \log\left(
        \operatorname{tr}_a e^{-K_{\mathrm{rec}}}
    \right)I_a
    \nonumber\\
    &=
    \frac{\omega}{2}\sigma_z
    +
    \log\left[
        2\cosh\left(\frac{\omega}{2}\right)
    \right]I_a.
    \label{eq:tn_normalized_generator}
\end{align}
The reconstructed state is
\begin{equation}
    e^{-K_a^{\mathrm{rec}}}
    =
    \begin{pmatrix}
        \dfrac{1}{1+e^\omega}&0\\[6pt]
        0&\dfrac{e^\omega}{1+e^\omega}
    \end{pmatrix}
    =
    \rho_a.
    \label{eq:tn_reconstructed_state}
\end{equation}
Hence
\begin{equation}
    K_a^{\mathrm{rec}}=K_a,
    \qquad
    K|_{\mathcal A}
    =
    K_a^{\mathrm{rec}}\otimes I_{\bar a}.
    \label{eq:tn_reconstructed_logical_modular}
\end{equation}
The conditional modular Hamiltonian has
therefore been recovered exactly.

\subsection{Recovery of the discrete RT area}
\label{subsec:tn_area_recovery}

Pulling the boundary modular Hamiltonian back
to the logical code gives
\begin{equation}
    V^\dagger(K_A\otimes I_{\bar A})V
    =
    K_a\otimes I_{\bar a}
    +
    (\log D)I_{\mathrm{code}}.
    \label{eq:tn_compressed_boundary_modular}
\end{equation}
Applying the reconstruction prescription,
\begin{align}
    \mathcal{L}_A^{\mathrm{rec}}
    &=
    V^\dagger(K_A\otimes I_{\bar A})V
    -
    K|_{\mathcal A}
    \nonumber\\
    &=
    (\log D)I_{\mathrm{code}}
    =
    \mathcal{L}_A^{\mathrm{known}}.
    \label{eq:tn_recovered_area}
\end{align}
The area operator obtained from the reconstructed
logical generator therefore agrees with the
independent bond-counting result.

As a numerical instance, take $q=1/5$ and $D=3$.
The boundary density matrix has eigenvalues
$1/15$ and $4/15$, each with multiplicity three.
The Lanczos calculation gives
\begin{equation}
    b_1=\log4,
    \qquad
    K_{\mathrm{rec}}
    =
    \begin{pmatrix}
        \log2&0\\
        0&-\log2
    \end{pmatrix}.
\end{equation}
Normalization yields
\begin{equation}
    K_a^{\mathrm{rec}}
    =
    \begin{pmatrix}
        \log5&0\\
        0&\log(5/4)
    \end{pmatrix}.
\end{equation}
The two logical eigenvalues of the modular
difference are
\begin{equation}
    \log15-\log5=\log3,
\end{equation}
and
\begin{equation}
    \log(15/4)-\log(5/4)=\log3.
\end{equation}
Thus
\begin{equation}
    \mathcal{L}_A^{\mathrm{rec}}
    =
    \log3\,I_{\mathrm{code}}.
    \label{eq:tn_numerical_area}
\end{equation}

This benchmark illustrates the reconstruction
starting from boundary modular information,
without supplying the logical modular
Hamiltonian as an input to the reconstruction
step. The explicit encoding and state
specify the test model and provide an
independent expression against which
the reconstructed result can be checked.
The labeled modular commutators determine
the traceless logical generator, its
scalar ambiguity is fixed by normalization,
and subtraction from the absolute boundary
modular generator on the code recovers
$\log D\,I_{\mathrm{code}}$.

At fixed $q$, varying $D$ leaves the
protected modular dynamics, Lanczos
coefficients, and Krylov complexity
unchanged, while the central area
contribution is $\log D$. The example
therefore exhibits both ingredients
of the proposal: direct access to
protected dynamics without projector
insertions, and recovery of the central
term through absolute modular information
that is absent from those dynamics.



\section{Existence of Islands?}\label{island}

In semiclassical models of an evaporating
black hole coupled to a nongravitating bath,
the island prescription reproduces a Page
curve through a change in the dominant
generalized-entropy saddle. An initially
dominant no-island saddle can be replaced
by an island saddle at the Page transition
\cite{Penington:2019npb,Almheiri:2019psf}.
The location of the corresponding QES and
the competition between saddles depend on
the model and require a gravitational
analysis. Here we discuss a conditional
application of the modular reconstruction
to the associated area contribution,
rather than an independent derivation
of island formation.

Determining an island’s location requires a full extremization in the gravitational bulk and depends sensitively on the geometry behind the horizon. This makes the problem especially difficult for one-sided or evaporating black holes, where the interior is dynamical and lacks a fixed, eternal description. Without a complete non-perturbative understanding of the bulk, the precise region encoding the information cannot be sharply defined. Thus, while the island formula explains what occurs, identifying where and how islands arise in general remains a challenge in quantum gravity.
In this section, we discuss a possible application of the
reconstruction developed in Secs.~ \ref{sec:area_modular_krylov-data} and \ref{sec:multisector_reconstruction} to the area contribution
associated with an island boundary. This interpretation
requires an appropriate identification of the protected
CFT algebra and its complementary entanglement wedge;
the reconstruction alone does not establish the existence
or dominance of an island saddle.

In order to study the evaporating black hole, we couple the CFT to a bath. The bath can be another CFT or a collection of qubits. Consider a general entangled state of the boundary as 
\begin{equation}
    \ket{\Psi_{bdy}} = \sum_i c_i \ket{\psi_{i,CFT}} \ket{\Tilde{i}}
\end{equation}
where $ \ket{\psi_{i,CFT}}$ are orthonormal states in the original CFT and $\Tilde{i}$ are states in the bath.
Let us consider $ \mathcal{A}_{code} = \mathcal{A}_{cg} \otimes \mathcal{B}$
where $ \mathcal{A}_{cg}$ is the coarse-grained algebra of the CFT dual to the algebra of the exterior of the black hole, and $ \mathcal{B} = \mathcal{L} (\mathcal{H}_{bath})$. The code subspace is the set of states constructed by acting $ \mathcal{A}_{code}$ on $ \ket{\Psi_{bdy}}$. The corresponding subspace has the structure of a Hilbert space that can be made via GNS construction 
\begin{equation}
    \mathcal{H}_{code}= \mathcal{H}_{bdy, GNS} \cong \mathcal{A}_{code} \ket{\Psi_{bdy}}
\end{equation}
Here, instead of dividing the boundary of the region $A$ and its complementary part, we divide the boundary into two parts: the entire CFT and the bath. The entire CFT plays the role of the region $A$ for us here. 
Let us define the set of time-evolved states of the boundary as
\begin{equation}
    \ket{\Psi_{bdy}(T)} = e^{-iT H_{bdy} }~  \ket{\Psi_{bdy}}.
\end{equation}
where $T$ is the real boundary time.
 For the "region" $A$ taken as the entire CFT (with the bath as complement), we compute the time-dependent modular Hamiltonian 
\begin{equation}
    K_{CFT} (T)= - \log \tr_{bath} \ket{\Psi_{bdy}(T)} \bra{\Psi_{bdy}(T)}
\end{equation}
At each physical time $T$, suppose that
$\mathcal A_{\mathrm{cg}}(T)$ is identified with the
protected algebra of the CFT entanglement wedge and that
its boundary representatives satisfy the modular-flow
conditions of Secs.~\ref{sec:area_modular_krylov-data} and \ref{sec:multisector_reconstruction}. Applying the reconstruction to a
sufficiently informative family of operators in this
algebra, together with the required normalization,
sector probabilities when applicable, and absolute
boundary modular information, then determines the
corresponding central contribution $\mathcal{L}_{\mathrm{CFT}}(T)$.
Its identification with a geometric area requires the
semiclassical interpretation of the modular relation.

We further assume that the semiclassical
description admits complementary CFT and
bath wedges with a common boundary.
Under this assumption, we identify the
CFT QES with the island boundary.
Equality of CFT and bath entropies for
a pure joint state does not by itself
establish this geometric identification.

This allows us to determine whether the QES is trivial or nontrivial and to identify the time at which the phase transition between these configurations possibly occurs.
This can manifest as a sharp change in the reconstructed area operator
corresponding to the new area term entering the center of the algebra. The transition time and the existence of the island are thus diagnosed directly from the boundary modular dynamics, without ever solving a bulk extremization problem.

Under these assumptions, the reconstructed
central term provides a candidate area
observable whose time dependence can be
compared with an independently established
island transition.
The proposal
offers a conditional boundary
probe of the island area, while establishing
island formation requires additional
control of the code and the competing
gravitational saddles.

\section{Discussion}

We have shown that the area operator of quantum
extremal surfaces can be reconstructed from
boundary modular data using OAQEC and
operator-resolved modular Krylov methods,
conditional on a specified code and protected
algebra. By applying the Lanczos algorithm to
modular-evolved operators and retaining
sufficient operator-resolved information, the
logical modular generator can be reconstructed
up to sector-dependent constants. Normalization
and sector probabilities obtained independently of the commutator data, for example through boundary central-projector measurements
fix these constants, allowing identification
of the area operator by subtraction from the
compressed boundary modular Hamiltonian,
without an additional bulk extremization
within the specified code. This framework
accommodates states supported on multiple
semiclassical geometry sectors and provides
a possible approach to studying island
formation and the Page transition in
evaporating black holes, subject to the
additional assumptions discussed in
Sec.~\ref{island}.
Our exact-code benchmark verifies the
reconstruction in a simple setting.

The motivation for this construction extends beyond whether the area operator can be reconstructed by other methods. Entanglement provides a foundational connection between boundary quantum information and bulk geometry, but entanglement entropy alone does not capture all aspects of gravitational dynamics, most notably the continued growth of black hole interiors after entropy saturation. This has motivated complexity-based descriptions of geometry, including the complexity–volume and complexity–action conjectures. These proposals associate boundary complexity with bulk geometric quantities, but do not by themselves supply a general boundary reconstruction procedure. Our work explores a complementary direction: within a specified operator-algebra quantum error-correcting code, operator-resolved modular Krylov data, combined with central-sector probabilities and absolute boundary modular information, provide a constructive route to the central area operator. The existence of alternative reconstruction methods does not diminish this connection; rather, it clarifies that our objective is to establish a modular-dynamical implementation of geometric reconstruction, not to demonstrate that complexity is indispensable. The result does not establish the complexity–volume or complexity–action conjectures, nor does it recover the area from the Krylov complexity function alone. It instead provides a concrete setting in which the operator dynamics underlying modular Krylov complexity participate in the reconstruction of a geometric observable.

One important point is that the JLMS relation
is understood after compression to the code
subspace, as written in \eqref{JLMSJLMS}.
For a state in $\mathcal H_{\mathrm{code}}$,
$K_A\otimes I_{\bar A}$ and its compression
agree on the component of their action
that remains in the code. However, only
the compressed generator necessarily
generates evolution within the code.
Defining
\begin{equation}
    K^{\mathrm{EFT}}
    =
    \left.
    P_{\mathrm{code}}
    (K_A\otimes I_{\bar A})
    P_{\mathrm{code}}
    \right|_{\mathcal H_{\mathrm{code}}},
\end{equation}
the component mapping code states outside
the code is
\begin{equation}
    (I-P_{\mathrm{code}})
    (K_A\otimes I_{\bar A})
    P_{\mathrm{code}}.
\end{equation}
The compressed JLMS identity alone does
not bound this component or establish
the modular-flow intertwining used in
the reconstruction. In a semiclassical
holographic code, the size of such
corrections must be controlled separately;
we do not assume a universal $O(1/N)$
leakage estimate.

Moreover, it would be interesting to relate
our boundary modular diagnostics to continuous
tensor network descriptions of CFT--bath
dynamics, which provide an explicit real-time
representation of information flow in such
systems \cite{Vardian:2024jll}. Our results
show how boundary modular data, together
with the specified code structure, can
probe the bulk area term, offering a
possible tool for studying black hole
interiors in time-dependent settings.

\section*{Acknowledgements}

I  would like to thank M. Alishahiha, H. Arfaei,  A. Faraji Astaneh, and K. Papadodimas for useful discussions and their valuable comments on the draft.
This work was supported by the "Tarh-e-Sobat" Grant No.  11/157663. 
The author acknowledges the use of OpenAI’s ChatGPT for assistance with manuscript editing
and technical discussions that helped clarify
the presentation of the
reconstruction framework. 

\bibliographystyle{apsrev4-2} 
\bibliography{refs.bib} 

\clearpage
\onecolumngrid

\appendix
\section{OAQEC and Area Operator}\label{App1}

OAQEC provides the natural language for entanglement wedge reconstruction in holography \cite{dong2016reconstruction}. A holographic code is specified by a code subspace 
$ \mathcal{H}_{code} \subset \mathcal{H}$ and encoding and recovery channels  $ \mathcal{E}, \mathcal{R}$.
In the Heisenberg picture, correctability implies that a protected set of observables is preserved on the code subspace,
\begin{equation}\label{OA}
    P_{code} ( \mathcal{R}\circ \mathcal{E})^*(O)P_{code} = P_{code} ~O~ P_{code}
\end{equation}
In OAQEC, this condition is imposed only for a von Neumann algebra 
$\mathcal{A} \subset \mathcal{A}_{code} = \mathcal{L} (\mathcal{H}_{code})$,
rather than for all operators. Information is then encoded redundantly according to the algebraic structure of 
$\mathcal{A}$ \cite{kribs2005unified, beny2007generalization}.

In finite dimensions, consider the decomposition of the Hilbert space as
\begin{equation}\label{decomposition}
     \mathcal{H} = \mathcal{H}_A \otimes \mathcal{H} _{\bar{A}}.
\end{equation}
For a generic von-Neumann algebra $ \mathcal{A}$, there always exists a block decomposition of the code subspace as 
\begin{equation}
    \mathcal{H}_{code} = \oplus_\alpha( \mathcal{H}_{A_1^\alpha}\otimes  \mathcal{H}_{\bar{A}_1^\alpha}\ )
\end{equation}
where
\begin{equation}
    \begin{aligned}
      & \mathcal{A} =  \oplus_\alpha( \mathcal{L}(\mathcal{H}_{A_1^a})\otimes  I_{\bar{A}_1^\alpha}\ )
        \\
      &\mathcal{A}' =  \oplus_\alpha( I_{A_1^\alpha} \otimes \mathcal{L}(\mathcal{H}_{\bar{A}_1^a}) \ ) 
    \end{aligned}
\end{equation}
while $ \mathcal{A}'$ is the commutant of $\mathcal{A}$. The center of the algebra labels superselection sectors $ \alpha$
\begin{equation}
     \mathcal{Z}_{\mathcal{A}}= \oplus_\alpha( \lambda_\alpha~ I_{A_1^\alpha} \otimes  I_{\bar{A}_1^\alpha}\ ). 
\end{equation}
Operators in 
$\mathcal{A}$ admit representatives acting only on the boundary region 
$A$, while operators in the commutant admit representatives on the complement, realizing complementary recovery as expected in AdS/CFT.
It is good to note that one possible choice of such operators can be given via the Petz map \cite{Cotler:2017erl, Bahiru:2022ukn, Vardian:2023fce} or some other appeoaches \cite{Faulkner:2017vdd, Faulkner:2018faa, Chandrasekaran:2021tkb, Chandrasekaran:2022qmq}, which is an operator constructed in the code subspace.

A central result of OAQEC is an algebraic formulation of the Ryu–Takayanagi relation \cite{Kamal:2019skn} as:

 In the case of the EWR, when the bulk algebra $ \mathcal{A} \subset \mathcal{L}(\mathcal{H}_{code})$ in the EW can be reconstructed from a boundary algebra $\mathcal{M}_A = \mathcal{L} (\mathcal{H}_A)$ and the commutant $ \mathcal{A}'$
from $\mathcal{M}_A'$, the algebraic entropy of the boundary algebra $\mathcal{M}_A$ satisfies an algebraic Ryu–Takayanagi formula
\begin{equation}\label{RT}
    S(\rho_{code}, \mathcal{M}_A) = \tr ( \rho_{code}~ \mathcal{L}_ A) +  S(\rho_{code}, \mathcal{A})
\end{equation}
where $  S(\rho_{code}, \mathcal{N})$ is the algebraic entropy \cite{Casini:2013rba} of the algebra $ \mathcal{N}$ and $ \mathcal{L}_A$ is the linear operator which is related to the area of the QES in holographic theories.

Now, let us briefly see how \eqref{RT} can be obtained from the OAQEC \cite{Harlow:2016vwg, Kamal:2019skn}.

In the context of holography, one can think of $ \mathcal{H}$ as the boundary Hilbert space and $ \mathcal{H}_{code}$ as the image of the bulk effective field theory under the isometry of embedding $V$. We always consider the subalgebra $ \mathcal{A} \subset \mathcal{A}_{code}= \mathcal{L}(\mathcal{H}_{code})$ to be a von-Neumann algebra and then 
\begin{equation}
    \mathcal{H}^{bulk}_{code}  = \oplus_\alpha ( \mathcal{H}_{a_\alpha} \otimes \mathcal{H}_{\bar{a}_\alpha})
\end{equation}
where $\mathcal{H}_{a_\alpha} \cong  \mathcal{H}_{A_1^\alpha}$ and $\mathcal{H}_{\bar{a}_\alpha} \cong  \mathcal{H}_{\bar{A}_1^\alpha}$.

In AdS/CFT, if we suppose that $ \mathcal{H}_A$ in \eqref{decomposition} corresponds to the Hilbert space of a spatial region $A$ on a boundary Cauchy slice, then $ \mathcal{H}_{a_{\alpha}}, \forall \alpha$ belongs to the operator algebra supported on the entanglement wedge of the region $A$, $ a = \mathcal{E}_A$.

If we introduce orthogonal bases 
\begin{equation}
    \mathcal{H}_{a_\alpha} = \text{span}\{ \ket{\alpha, i}_a\}, ~~~ \mathcal{H}_{\bar{a}_\alpha} = \text{span}\{ \ket{\alpha, j}_{\bar{a}}\}
\end{equation}
then, the bulk code subspace bases are 
$ \ket{\alpha, ij}_{bulk}= \ket{\alpha, i}_a\otimes \ket{\alpha, j}_{\bar{a}} $.
In \cite{Harlow:2016vwg}, it is proved that 
there exist
unitaries $ U_A$ and $U_{\bar{A}}$ such that 
\begin{equation}
    \ket{\alpha, ij}_{bulk} \longrightarrow~ \ket{\alpha, ij}_{code} = U_A U_{\bar{A}}\Big(\ket{\alpha, i}_{A_1^\alpha} \ket{\alpha, j}_{\bar{A}_1^\alpha} \ket{\chi_\alpha}_{A_2^\alpha \bar{A}_2^\alpha} \Big)
\end{equation}
where $ \mathcal{H}_{code} = \text{span}\{ \ket{\alpha, i,j}_{code}\}$,
and
\begin{equation}
  \ket{\alpha, i}_a \cong \ket{\alpha, i}_{A_1^\alpha}, ~~~  \ket{\alpha, j}_{\bar{a}} \cong \ket{\alpha, j}_{\bar{A}_1^\alpha}
\end{equation}
The unitaries are the operators that encode the logical state by mixing it with redundant degrees of freedom $ \chi_\alpha$.
The $\chi_\alpha $ state must correspond to the high-energy quantum gravity degrees of freedom that were integrated out to define the effective field theory of the gravity theory. 
For an arbitrary encoded state $ \rho_{code}$ we have 
$ \rho \Big|_{\mathcal{M}_A} = \rho_A \otimes \frac{I_{\bar{A}_2^\alpha}}{\bar{A}_2^\alpha}$ \cite{Harlow:2016vwg}, where
\begin{equation}\label{rho}
    \rho_A = \tr_{\bar{A}} \rho_{code}  = U_A \big( \oplus_\alpha ( p_\alpha ~\rho_{A_1^\alpha} \otimes \chi _{A_2^\alpha})\big)U_A^\dagger
\end{equation}
while $\chi _{A_2^\alpha} = \tr_ {\bar{A}_2^\alpha} \ket{\chi_\alpha} \bra{\chi_\alpha}$ and 
$ \rho_{A_1^\alpha }$ act on $ \mathcal{H}_{A_1^\alpha}$ in the same way that $ \rho _{a_\alpha}$ do on $ \mathcal{H}_{a_\alpha}$.
Finally, if we define
\begin{equation}\label{area}
    \mathcal{L}_A = \mathcal{L}_{\bar{A}} = \oplus_\alpha S(\chi_\alpha, \mathcal{M}_A) I_{a_\alpha} \otimes I_{\bar{a}_\alpha},
\end{equation}
we arrive at the algebraic Ryu-Takayanagi formula \eqref{RT}.
Now, in the context of entanglement wedge reconstruction in holography, comparing von-Neumann with the Ryu-Takayanagi formula in the gravitational theory with the Einstein-Hilbert action, the area operator given by
\begin{equation}\label{area2}
    \mathcal{L}_A = \frac{\hat{\mathcal{A}}_X}{4G_N}= \oplus _\alpha S(\chi_\alpha, \mathcal{M}_A) I_{a_\alpha} \otimes I_{\bar{a}_\alpha}
\end{equation}
Therefore, the area operator must be in the center of representation of the algebra $ \mathcal{\mathcal{A}}$ associated with the algebra in the $ a= \mathcal{E}_A$.

In the case that $\mathcal{A}$ is not a factor, in the bulk effective field theory it is related to the presence of the gauge symmetry \cite{Harlow:2016vwg, Akers:2018fow} and therefore, the area operator is part of the "edge mode" /"soft hair" \cite{Harlow:2016vwg}.
In addition to the gauge symmetry, we can have a non-trivial center in the case that we have a superposition of the semi-classical states. For example, empty AdS and some field excitation around empty AdS share the same semi-classical background and thus, they can be in one code subspace, but they might be in different $\alpha$-blocks since $ \frac{\hat{\mathcal{A}}_X}{4G_N}$ differs for them. Therefore, considering the superposition of the semi-classical states, one can generalize the RT formula to 
\begin{equation}
    \mathcal{\mathcal{L}_A} = \frac{\hat{\mathcal{A}}_X}{4G_N} = \sum _\alpha \frac{\hat{A}(X_{min}^\alpha)}{4G} P_\alpha + O(1/N)
\end{equation}
where $P_\alpha$ is the projection onto the subset of states in the code subspace with the same background geometry and $ X_{min}^\alpha$ is QES on that geometry. The off-diagonal terms are discussed in detail in \cite{Almheiri:2016blp}.


\section{Review of Krylov complexity}\label{krylov}
\subsection{Krylov Complexity of states and operators}

In \cite{balasubramanian2022quantum}, a concept known as Krylov state complexity is introduced. Krylov complexity has been defined both for states and for operators, and more recently generalized to modular Krylov complexity \cite{Caputa:2023vyr}, where the evolution is generated by a modular Hamiltonian rather than the physical Hamiltonian, and multi-generator unitary evolution \cite{FarajiAstaneh:2025thi}.
Krylov complexity provides an intuitive visualization of a wavefunction dispersing throughout the Hilbert space, which is basis independent. It is also referred to as "spread complexity".
It measures how far the target state spreads in the Hilbert space.

Consider a quantum system with a time-independent Hamiltonian $H$. A state $ \ket{\psi (t)}$
is time-evolved under the Schrodinger equation
has a formal power series expansion 
\begin{equation}
    \ket{\psi(t)} = \sum _{n=0}^\infty \frac{(-it)^n}{n!} \ket{\psi_n}
\end{equation}
while $ \ket{\psi_n} = H^n \ket{\psi_0}$. Therefore, the time-evolved state  is a linear combination of 
\begin{equation}\label{basis}
\{ \ket{\psi_0}, \ket{ \psi_1} = H \ket{\psi_0},\ket{\psi_2} = H^2\ket{\psi_0},~ ... \} 
\end{equation}
The subspace $ \mathcal{H}_{\psi}$ which is spanned by this set of vectors, is called "Krylov subspace". Notice that in general, this basis is not orthogonal.
The Gram-Schmidt procedure applied to $ \ket{\psi_n}$ generates an orthogonal basis 
\begin{equation}
    \mathcal{K}_\psi= \{ \ket{K_n}: n=0,1,2,..., d_\psi-1\}
\end{equation}
where we define $ d_\psi = \dim \mathcal{H}_\psi$ for the subspace of the full Hilbert space explored by the evolution of $ \ket{\psi_0} = \ket{K_0}$. In general, this subspace can be of infinite dimension.

Using the ordinary inner product, one can orthogonalize the basis \eqref{basis} through the "Lanczos algorithm" \cite{viswanath1994recursion, gordon1968error}, which leads to two sets of  Lanczos coefficients $ \{a_n, ~b_n \}$,  which
encode all information regarding the system’s dynamics.
In the case that $ d_\psi$ is finite, the Lanczos algorithm will end at some point where $ b_{n\geq d_\psi} =0$.

These coefficients correspond to the elements of the Hamiltonian in a basis that is tridiagonal, since
\begin{equation}\label{22}
H \ket{K_n} = a_n \ket{K_n} + b_{n+1}\ket{K_{n+1}} + b_n \ket{K_{n-1}}.
\end{equation}
We can expand the time-evolved state in terms of the Krylov basis as 
\begin{equation}
    \ket{\psi(t)} = \sum _{n=0}^{K_\psi -1} \phi_n(t) \ket{K_n}
\end{equation}
while $ \sum |\phi_n(t)|^2 =1$ and $ \phi_n(t) = \langle K_n\ket{\psi(t)}$.
The Krylov state complexity or spread complexity of the state $ \ket{\psi (t)}$ is defined as 
\begin{equation}
    C_k (t) = \sum _{n=0} ^{K_\psi -1} n | \phi_n(t)|^2 = \bra{\psi(t)} \sum _{n=0}^{d_\psi} n \ket{K_n}\bra{K_n} \psi(t) \rangle
\end{equation}
which measures the "spreading" of the initial state along the one-dimensional chain.

Following the same approach for quantum states, one can define Krylov complexity for the operators by considering the evolution of the operator with a time-independent Hamiltonian in the Heisenberg Picture. 
One can create the Krylov basis for a given operator in terms of the nested commutator with the Hamiltonian
\begin{equation}
    O(t)= \sum_{n=0}^\infty \frac{(-it)^n}{n!} \mathcal{L}^n O_0 = \sum _{n=0} ^{d_O-1} i^n \psi_n(t)~ O_n
\end{equation}
where $ \mathcal{L} := [H,.]$ is the Liouvillian operator, and we refer to the Krylov space of the operator as 
$ \mathcal{K}_O = \{O_n: n= 0,1,..., d_O-1\}$. In order to apply the Lanczos algorithm, one also needs to introduce a notion of an inner product.

\subsection{Lanczos coefficients and Moment expansion}\label{algorithm}

We review how the Lanczos coefficients can be calculated from the return amplitude 
\begin{equation}\label{RA}
    R(t) = \langle \psi(0) | \psi (t)\rangle 
\end{equation}
for a given initial state $ \ket{\psi(0)}$. We can compute the set of $ \{a_n\} , \{b_n\}$ by the following moment method.

By using the Taylor expansion of the return amplitude 
\begin{equation}
  R(t) = \sum _{n=0} M_n \frac{(-it)^n}{n!}  
\end{equation}
the moments $M_n$ defined as 
\begin{equation}\label{2}
M_n = \frac{1}{(-i)^n} \frac{d^n R(t)}{dt^n}\Big|_{t=0}.
\end{equation}
One can compute the moments from the Fourier transform of the return amplitude as
\begin{equation}
    M_n =\int _{- \infty}^ {\infty} \frac{d\omega}{2\pi} \omega ^n f(\omega)
\end{equation}
while
\begin{equation}
    f(\omega) = \int _{- \infty}^ {\infty} dt e^{i\omega t } R(t).
\end{equation}
Note that if $ f(\omega)$ is an even function $M_{2n+1}= 0$ which implies that $ a_n =0$.

There is an algorithm to compute the Lanczos coefficients from the moments \cite{viswanath1994recursion, gordon1968error}.
The relation between the moments and these sets of coefficients are most conveniently expressed in terms of two arrays of auxiliary quantities $ L_k ^{(n)} $ and  $ M_k ^{(n)} $.

Given a set of moments $ M_0 \equiv 1$, $ M_1,..., M_{2K+1}$ the coefficients $ a_0,..., a_K$ and $ b_1, ... ,b_K$ are obtained by initializing 
\begin{equation}
    M^{(0)} = (-1)^k M_k, ~~~~~ L_k^{(0)} = (-1)^{k+1} M_{k+1}
\end{equation}
for $ k=0,..., 2K$ and then applying the recursion relations
\begin{equation}\label{ab}
    \begin{split}
        M_k^{(n)}& = L_k^{(n-1)} - L_{n-1}^{(n-1)} \frac{ M_k^{(n-1)}}{M_{n-1}^{(n-1)}}
        \\
        L_k^{(n)} & = \frac{M_{k+1}^{n}}{M_n^{(n)}} - \frac{M_k^{(n-1)}}{M_{n-1}^{(n-1)}}
    \end{split}
\end{equation}
for $ k= n,..., 2K-n+1$ and $ n=1,..., 2K$. The resulting coefficients are
\begin{equation}\label{33333}
    b_n = \sqrt{ M_n^{(n)}}, ~~~~~ a_n = - L_n ^{(n)}, ~~~~~ n=0,...K.
\end{equation}

\subsection{Modular Krylov Complexity}

Modular Krylov complexity extends the usual Krylov construction by replacing the physical Hamiltonian with a modular Hamiltonian \cite{Caputa:2023vyr}. Given a (reduced) density matrix, the modular Hamiltonian is defined as 
\begin{equation}
    K = - \log \rho.
\end{equation}
Consider the decomposition of the total Hilbert space $\mathcal{H}$ into subsystems $A$ and its complement $\Bar{A}$. We have 
$ \mathcal{H} = \mathcal{H}_A\otimes \mathcal{H}_{\Bar{A}} $. Given an initial state $ \ket{\psi_0} \in \cal H$, it has a Schmidt decomposition as 
\begin{equation}\label{initialstate}
    \ket{\psi_0} = \sum _i \sqrt{\lambda_i} \ket{i}_A \ket{i}_{\Bar{A}}.
\end{equation}
The reduced density matrix of subsystem $A$ is
\begin{equation}
    \rho_A = \Tr_{\Bar{A}}\big(\ket{\psi_0} \bra{\psi_0}\big) := e^{-K_A}
\end{equation}
where $K_A$ is the "modular Hamiltonian" of the subsystem $A$.
In \cite{Caputa:2023vyr}, they generate modular time evolution. For a reference state $ \ket{\psi_0}$ in the support of $\rho_A$, one can define 
\begin{equation}
    \ket{\psi (s)} = e^{-is~ K_A\otimes I_{\Bar{A}}} ~\ket{\psi_0}
\end{equation}
where $s$ is the "modular time". 
It is important to note that the total modular Hamiltonian of the state is $ K= K_A \otimes I_{\bar{A}} -I_A \otimes K_{\Bar{A}}$ where $ K_{\Bar{A}} = - \log \rho_{\Bar{A}} $ and $ \ket{\psi_0}$ is invariant under evolution with $K$. But evolution with $ K_A \otimes I$   leads to non-trivial evolution of the initial state

As in the Hamiltonian case, repeated action of $ K_A\otimes I_{\Bar{A}}$ on $\ket{\psi_0}$ generates the modular Krylov subspace 
\begin{equation}
    \mathcal{H}_\psi^{(mod)} = \text{span} \{ \ket{\psi_0},  (K_A\otimes I_{\Bar{A}})  \ket{\psi_0}, ( K_A\otimes I_{\Bar{A}})^2  \ket{\psi_0},... \}.
\end{equation}
Applying the Lanczos algorithm to the sequence of $ \{ \ket{\psi_0},   \ket{\psi_1^{(mod)}}= K_A\otimes I_{\Bar{A}}  \ket{\psi_0}, ... \}$ yields two sets of Lanczos coefficients $ \{ a_n, b_n\}$ and  an orthonormal "modular Krylov basis" as
\begin{equation}
       \mathcal{H}_\psi^{(mod)} = \text{span} \{ \ket{K_n ^{(mod)}}, n=0,1,2,..., d_{\psi} ^{(mod)}-1 \}.
\end{equation}
where $ \dim  \mathcal{H}_\psi^{(mod)} = d_{\psi} ^{(mod)}$. The modular evolved state can be decomposed in this basis as 
\begin{equation}
    \ket{\psi(s)} = \sum _{n=0}^{d_{\psi} ^{(mod)}-1 } ~ \phi_n(s)~ \ket{K_n ^{(mod)}},~~~ \sum_n |\phi_n(s)|^2=1
\end{equation}
and finally, the Krylov complexity of states is defined by 
\begin{equation}
    C_\psi^{(mod)}(s) = \sum _{n=0} ^{ d_{\psi} ^{(mod)}-1} n ~|\phi_n(s)|^2
\end{equation}
This quantity measures how strongly the state spreads along the one-dimensional chain generated by the modular Hamiltonian $ K_A \otimes I_{\Bar{A}}$, and thus probes the complexity of modular flow, rather than the ordinary real-time dynamics.

Practically, the Lanczos coefficients can be calculated from the modular return amplitude, defined as 
\begin{equation}\label{RA}
    R(s) = \langle \psi_0 | \psi _(s)\rangle 
\end{equation}
following moment method \cite{viswanath1994recursion, gordon1968error}.

\end{document}